\documentclass[12pt]{article}

\usepackage{amsfonts,amsmath}
\usepackage{amssymb}
\usepackage{ytableau}
\usepackage{booktabs}
\usepackage{mathtools} 
\usepackage{tikz-cd} 
\usepackage{comment}

\usepackage{pgfplots}
\pgfplotsset{compat=1.5}
\usepgfplotslibrary{fillbetween} 
\usepackage{tikz}
\usetikzlibrary{arrows.meta,bending,patterns,angles,calc}
\usetikzlibrary{patterns.meta}
\usepackage{siunitx}
\usetikzlibrary{decorations.shapes}
\tikzset{decorate sep/.style 2 args=
{decorate,decoration={shape backgrounds,shape=circle,shape size=#1,shape sep=#2}}}

\pgfplotsset{ignore zero/.style={%
  #1ticklabel={\ifdim\tick pt=0pt \else\pgfmathprintnumber{\tick}\fi}
}} 

\newcommand{\dr}{{{\rm d}}}

\makeatletter \@addtoreset{equation}{section} \makeatother

{\vspace{3mm} }

\def\al{\alpha}

\def\*{\star}
\def\e{\mathbf{e}}
\def\E2{\mathbf{E}}

\def\rmx{\mathrm{x}}

\newcommand{\be}{\begin{equation}}
\newcommand{\ee}{\end{equation}}
\newcommand{\bee}{\begin{eqnarray}}
\newcommand{\beee}{\begin{array}}
\newcommand{\eee}{\end{eqnarray}}
\newcommand{\eeee}{\end{array}}
\newcommand{\gb}{\beta}
\newcommand{\gga}{\gamma}

\newcommand{\gep}{\epsilon}

\newcommand{\go}{\omega}

\newcommand{\y}{{\bar{y}}}
\newcommand{\dal}{\dot \alpha}
\newcommand{\dgb}{\dot \beta}
\newcommand{\dgga}{\dot \gamma}

\newcommand{\p}{\partial}
\newcommand{\bp}{\bar\partial}

\newcommand{\ff}{\frac}
\newcommand{\rom}[1]{\uppercase

\expandafter{\romannumeral #1\relax}}

\begin{document}
    
\begin{flushright}
FIAN/TD/12-2026\\
\end{flushright}

\vspace{0.5cm}
\begin{center}
{\large\bf Self-dual higher spin theory: Poincar\'{e} invariance and new solutions.}

\vspace{1 cm}

\textbf{I.S.~Faliakhov$^{1, 2}$}\\

\vspace{1 cm}

\textbf{}\textbf{}\\
 \vspace{0.5cm}
 \textit{$^{1}$I.E. Tamm Department of Theoretical Physics,
Lebedev Physical Institute,}\\
 \textit{ Leninsky prospect 53, 119991, Moscow, Russia }\\
 \vspace{0.5cm}
 {\it
			$^2$Moscow Institute of Physics and Technology,\\
			Institutsky lane 9, 141700, Dolgoprudny, Moscow region, Russia}

\par\end{center}

\begin{center}
\vspace{0.6cm}
e-mail: faliakhov.is@phystech.edu \\
\par\end{center}

\vspace{0.4cm}

\begin{abstract}
\noindent In this paper we study the self-dual higher spin theory in $4d$ recently proposed in \cite{Didenko:2022qga}. The typical vacuum of higher spin theories is an empty AdS space-time. We consider solutions of special form: space-time geometry associated with spin $S=2$ field is an AdS, and other fields may have nonzero values, such that the global space-time symmetry of the vacuum is broken to $3d$ Poincar\'{e}. We show that the only field possessing this property is a two-parametric scalar vacuum. We also provide a new family of solutions that generalize this scalar vacuum. In addition to the scalar, the family consists of a fermion $S = 1/2$ in the bulk propagating along the radial direction in Poincar\'{e} coordinates and trivial fields of all spins. Fields are trivial in the sense that they do not have gauge fields in the bulk and correspond to trivially conserved currents (i.e. constants) on the conformal boundary.
\end{abstract}

\newpage
\tableofcontents
\newpage

\section{Introduction}

Higher spin (HS) gauge theories are aimed at describing interactions of the fields of arbitrary spins. There are a number of no-go results against interacting massless HS in flat space-time \cite{Weinberg:1964ew, Coleman:1967ad, Aragone:1979hx}. Progress was made possible by the transition to an AdS space-time \cite{Fradkin:1987ks}, \cite{Fradkin:1986qy}. The system describing interactions of massless gauge HS fields has been obtained by Vasiliev \cite{Vasiliev:1992av}. Vasiliev's equations are based on the HS algebra, the maximal space-time subalgebra of which corresponds to isometries of AdS. The simplest vacuum of the theory is an empty AdS space-time; see \cite{Vasiliev:1999ba, Bekaert:2004qos, Didenko:2014dwa} for reviews.

In what follows, we are working in $4d$. The HS algebra is constructed via the set of auxiliary variables $(y_{\alpha}, \y_{\dot\alpha})$ that are $sl(2,C)$ spinors. The construction uses the isomorphism $so(3,1) \sim sl(2,C)$ which appears to be helpful in $4d$. The equations schematically are of the form:
\begin{equation}
\begin{split}\label{equations}
    & \dr_x \omega = \sum_{n = 0}^{\infty} \mathcal{V}_n(\omega, \omega, C^n), \\
    & \dr_x C = \sum_{n = 1}^{\infty} \Upsilon_n(\omega, C^n).
\end{split}
\end{equation}
The fields are formal series in variables $(y_{\alpha}, \y_{\dot\alpha})$. Specifically, $\omega$ is a space-time $1$-form containing gauge fields for spin $S \geqslant 1$ and $C$ is a space-time $0$-form which consists of generalized Weyl tensors together with the scalar $S = 0$ and the fermion $S = 1/2$. The r.h.s. contains the so-called vertices, which describe interactions. Vertices must be constrained to accomplish simple requirements. First, equations \eqref{equations} must be consistent in the sense of $\dr_x^2 = 0$. Second, they must correctly describe the free-field case, i.e., they must reproduce the Fronsdal equations \cite{Fronsdal:1978rb, Fang:1978wz}. The four dimensional theory has one free complex parameter $\eta$, which enters the vertices in the following fashion:
\begin{equation}
\begin{split}
    & \mathcal{V}_n(\omega, \omega, C^n) = \sum_{i+j = n} \eta^i\bar\eta^j\cdot \mathcal{V}_{i,j}(\omega, \omega, C^n), \\
    & \Upsilon_n(\omega, C^n) = \sum_{i+j = n-1} \eta^i\bar\eta^j \cdot \Upsilon_{i,j}(\omega, C^n).
\end{split}
\end{equation}
One parameter from the pair $\eta, \bar\eta$ could be set to zero without ruining the consistency. Such a truncated sector is called (anti-)\textit{self-dual}. By setting $\eta = 0$ we cut out the following subset from the complete set of vertices:
\begin{equation}
    \mathcal{V}(\omega, \omega, C^n) = \sum_{j = n} \bar\eta^j \cdot \mathcal{V}_{0,j}(\omega, \omega, C^n), \quad \Upsilon(\omega, C^n) = \sum_{j = n-1} \bar\eta^j \cdot \Upsilon_{0,j}(\omega, C^n) \label{self-dual-vertices}.
\end{equation}
For each spin $S$, the corresponding $so(3,1)$ Weyl tensor decomposes into a pair of $sl(2,C)$ spin-tensors
\begin{equation*}
    C^{a(S), b(S)} \sim C^{\alpha(2S)} \oplus \bar{C}^{\dot\alpha(2S)}.
\end{equation*}
The self-duality condition in terms of $sl(2,C)$ tensors means that only one from the pair acts as a source for HS gauge fields. Such a truncation leads to a non-unitary theory. Nevertheless, the self-dual sector is consistent, it is simpler and can be studied independently. Unlike the full HS theory, the self-dual case admits the flat limit that was originally discovered by Metsaev \cite{Metsaev:1991mt}; see also \cite{Ponomarev:2017nrr}. 

In this paper, we study the generating system for the self-dual HS fields recently proposed by Didenko \cite{Didenko:2022qga}; see also \cite{Didenko:2023vna} for any $d$ offshell system. Vasiliev's equations contain interaction vertices through the differential equations in the auxiliary space. The solution has a natural freedom in field redefinition, which affects the form of these vertices. Understanding the proper field variables is a significant part of the HS (non-)locality problem \cite{Prokushkin:1998bq, Ponomarev:2017qab, Sleight:2017pcz, Vasiliev:2022med, Didenko:2022eso, Neiman:2023orj}. The main feature of Didenko's system is that it differs from the original realization of Vasiliev's so-called large HS algebra, which was first observed in \cite{Didenko:2019xzz} as a contraction. This construction of the algebra leads to a straightforward way of restoring the interaction vertices from the generating equations in a manifestly local form \cite{Didenko:2024zpd} (see also \cite{Sharapov:2022nps}). This makes the self-dual system an excellent toy model for analyzing properties of the HS systems. We also note that the connection between the system from \cite{Didenko:2022qga} and the holomorphic sector of Vasiliev's equations is not known to all orders; see \cite{Didenko:2026nag} for details and for an extension of \cite{Didenko:2022qga} to the full $4d$ interactions.

HS theories are based on infinite-dimensional gauge symmetry. There are expectations in the literature regarding this. Specifically, superstring theory is anticipated to represent a broken phase of a theory with HS-like symmetry. This idea originally came from an observation that the tensionless limit leads to a significant extension of the symmetries of the string \cite{Gross:1988ue}; see also \cite{Sundborg:2000wp, Bonelli:2003kh, Beisert:2004di} for related studies and \cite{Cho:2026mjp} for the very recent discussion. The spectrum of the string naturally contains fields of all spins; these states are massive and become massless in the tensionless limit, a picture reminiscent of the Higgs mechanism. 

The specific mechanism implementing this idea remains unknown. The simplest cases of HS lack the diversity of states found in superstrings. The rich extension of HS was recently proposed by Vasiliev in \cite{Vasiliev:2018zer}, with further analysis of free dynamics available in \cite{Tarusov:2025sre}. However, due to the technical complexity of these models, progress is still limited. Another interesting proposal is provided in \cite{Metsaev:1999kb} by Metsaev, where it is suggested that string theory might emerge on the conformal boundary of AdS in the broken phase of HS. 

One way or another, in order to study symmetry breaking, one has to have a vacuum of high residual symmetry. This is our starting point. In \cite{Didenko:2023txr}, it was shown that the unconstrained HS equations admit a vacuum scalar solution, which breaks the symmetry of AdS$_{d+1}$ down to a Poincar\'{e} in $d$. This is a two-parametric solution that contains conformal branches of dimensions $\Delta = 2$ and $\Delta = d - 2$. A similar vacuum exists on-shell in four dimensions. The generating equations in $4d$ are technically simpler, allowing one to study the symmetry-broken phase; see \cite{Didenko:2025xca} for this analysis.

In this paper, we look for all solutions of the $4d$ self-dual HS that break the symmetry of the AdS$_{3+1}$ space-time down to a Poincar\'{e} in $2+1$, provided the space-time geometry remains AdS. To this end, we use the global symmetry equations for a general exact solution, constrained by the presence of translations and Lorentz generators. We observe that this condition leaves no room for other solutions different from the earlier obtained two-parametric scalar vacuum. We also observe that this scalar solution can be generalized if one relaxes the Poincar\'{e} invariance condition, resulting in a new family of solutions. These are new, exact, and yet simple solutions. We argue that they correspond to a set of trivially conserved currents on the $3d$ conformal boundary. Each solution carries spin-$S$ degrees of freedom, although they do not have corresponding gauge fields in the bulk. Conformal dimension of the spin-$S$ field is $\Delta_S = S + 1$ which corresponds to a known unitarity boundary in dimension $d = 3$.

\paragraph{} The paper is structured as follows. Section 2 contains a review of the self-dual HS proposed in \cite{Didenko:2022qga}. In section 3, we consider the Poincar\'{e} invariant solutions, showing that they are the only ones obtained in \cite{Didenko:2025xca}. Section 4 provides new solutions of the self-dual HS that generalize the scalar vacuum. Finally, section 5 offers a discussion of the findings.

\section{Self-dual HS theory}

\paragraph{Unfolded equations.}
General HS theory in four dimensions is naturally formulated in the so-called \textit{unfolded form} \cite{Vasiliev:1988sa} (see for recent applications \cite{Misuna:2024ccj, Misuna:2024dlx, Misuna:2026bhy, Iazeolla:2025btr} and for a review \cite{Vasiliev:1999ba})
\begin{subequations} \label{unfoldedEqs}
\begin{align}
    & \dr_x \omega = \sum_{n = 0}^{\infty} \mathcal{V}_n(\omega, \omega, C^n),\label{unfoldedEqs1}\\
    & \dr_x C = \sum_{n = 1}^{\infty} \Upsilon_n(\omega, C^n).\label{unfoldedEqs2}
\end{align}
\end{subequations}
Here $\omega(x|Y)$ is a space-time $1$-form containing gauge fields and $C(x|Y)$ is a space-time $0$-form consisting of generalized Weyl tensors for all spins together with the spin $S = 0, 1/2$ fields. These are generating functions depending on space-time coordinates $x$ and auxiliary spinor variables $Y = (y_{\alpha}, \y_{\dot\alpha}),\ \alpha, \dot\alpha = 1,2$. Throughout the paper we denote by $\alpha(n)$ a group of $n$ symmetrized indices, and use the notation $y_{\alpha(n)} := \underbrace{y_{\alpha}...y_{\alpha}}_{n\ \text{times}}$.
\begin{equation*}
\begin{split}
    & \omega(x| y, \y) = \sum_{n,m} \frac{1}{n!m!} \omega(x)^{\alpha(n),\dot\beta(m)}y_{\alpha(n)}\y_{\dot\beta(m)}, \\
    & C(x| y, \y) = \sum_{n,m} \frac{1}{n!m!} C(x)^{\alpha(n),\dot\beta(m)}y_{\alpha(n)}\y_{\dot\beta(m)}.
\end{split}
\end{equation*}
Spinor indices are contracted via antisymmetric $\epsilon_{\alpha\beta},\ \epsilon_{\dot\alpha\dot\beta}$. We use the convention $\epsilon_{\alpha\beta}\epsilon^{\alpha\gamma} = \delta_{\beta}{}^{\gamma} = -\epsilon_{\beta\alpha}\epsilon^{\alpha\gamma}$ and the short notation for contractions $\epsilon^{\alpha\beta}u_{\alpha} v_{\beta} := uv = -vu $. The auxiliary variables $Y$ form an algebra governed by the \textit{star-product}:
\begin{equation}
    f(y, \y) \star g(y, \y) = \int\frac{d^2ud^2vd^2\bar ud^2\bar v}{(2\pi)^4} f(y + u, \y + \bar u) g(y + v, \y + \bar v) e^{iuv + i\bar u \bar v}. \label{starproduct}
\end{equation}
This is also known as Moyal product. The r.h.s. of \eqref{unfoldedEqs} are the so-called \textit{vertices}. Four dimensional theory admits a free complex parameter $\eta$, which enters vertices in the following way:
\begin{equation}
\begin{split}
    & \mathcal{V}_n(\omega, \omega, C^n) = \sum_{k = 0}^n \eta^k\bar \eta^{n-k} \cdot \mathcal{V}_{n,k}(\omega, \omega, C^n), \\
    & \Upsilon_n(\omega, C^n) = \sum_{k = 0}^{n-1}\eta^k \bar\eta^{n-k-1}\cdot \Upsilon_{n,k}(\omega, C^n).
\end{split}
\end{equation}
The first vertices are fixed as follows (see, e.g., \cite{Vasiliev:1999ba}):
\begin{equation}
    \mathcal{V}_0(\omega, \omega) = -\omega \star \omega,\ \Upsilon_{1}(\omega, C) = -\omega \star C + C \star \pi(\omega). \label{initialData}
\end{equation}
Here $\pi(\omega(y, \y)) = \omega(-y, \y)$ is an automorphism of our algebra, that provides the correct representation of $C$. Another vertex is constrained in a special way 
\begin{equation}
    \mathcal{V}_1(\omega_{AdS}, \omega_{AdS}, C) = \frac{\bar \eta}{2}\bold e_{\gamma}{}^{ \dot\alpha} \wedge \bold e^{\gamma\dot\beta}\bp_{\dot\alpha}\bp_{\dot\beta}C(0, \y) + \frac{\eta}{2}\bold e^{\alpha}{}_{ \dot\gamma} \wedge \bold e^{\beta\dot\gamma}\p_{\alpha}\p_{\beta}C(y, 0). \label{CentralOnMassShellTheorem}
\end{equation}
Here $\bold e^{\alpha\dot\beta}$ is a veirbein of the $4d$ AdS. The special choice of vertices \eqref{initialData}, \eqref{CentralOnMassShellTheorem} ensures that the unfolded system reproduces Fronsdal equations in the AdS background \cite{Fronsdal:1978rb}, \cite{Fang:1978wz}, \cite{Vasiliev:1992av}. Specifically, one should fix an exact solution of an empty AdS, then linearize equations \eqref{unfoldedEqs} for $(\omega_{AdS} + \omega, C)$ and get the unfolded form of the Fronsdal equations. This result is known as a central on-mass-shell theorem:
\begin{equation}\label{CentralOnMassShellTheoremEqs}
\begin{split}
    & \dr_x \omega + \{\omega_{AdS}, \omega\}_{\star} = \frac{\bar\eta}{2}\bold e_{\gamma}{}^{ \dot\alpha} \wedge \bold e^{\gamma\dot\beta}\bp_{\dot\alpha}\bp_{\dot\beta}C(0, \y) + \frac{\eta}{2}\bold e^{\alpha}{}_{ \dot\gamma} \wedge \bold e^{\beta\dot\gamma}\p_{\alpha}\p_{\beta}C(y, 0), \\
    & \dr_x C + \omega_{AdS} \star C - C \star \pi(\omega_{AdS}) = 0.
\end{split}
\end{equation}
To obtain the self-dual sector, one has to set $\eta = 0$. The truncation appears to be consistent \cite{Iazeolla:2007wt}. The spin $S$ modules are singled out by the following operators
\begin{subequations}\label{prime}
	\begin{align}
		&\left(y^{\al}\frac{\p}{\p y^{\al}}+\bar y^{\dal}\frac{\p}{\p\bar y^{\dal}}\right)\go=2(S-1)\go\,,\label{Omega:phys}\\
		&\left(y^{\al}\frac{\p}{\p y^{\al}}-\bar y^{\dal}\frac{\p}{\p\bar y^{\dal}}\right)C=2s\, C\,,\label{C:phys}
	\end{align}
\end{subequations}
where $s=\pm S$ denotes helicity. The spin $S$ module contains \textit{primary} and \textit{descendant} fields. The descendants are expressed via derivatives of primaries on-shell. Primaries are those that cannot be expressed via derivatives of other fields. Thus, equations \eqref{CentralOnMassShellTheoremEqs} tells that HS gauge fields, that are components of $\omega$, are sourced by $C^{\alpha(n)}$ and $\bar C^{\dot\alpha(n)}$. Whereas in the (anti-)self-dual sector, only one of the pair source the gauge fields.

\paragraph{Generating system.} All possible interaction vertices in the r.h.s. of \eqref{unfoldedEqs} can be found using Vasiliev's generating system approach. In this paper, we focus on the self-dual HS generating equations, which were proposed in \cite{Didenko:2022qga}.

Equations are formulated using a new spinor variable $z_{\alpha} = (z_1, z_2)$ which expands $Y$-space. We call this new algebra \textit{enlarged} HS algebra. The star-product should be enlarged too:
\be\label{limst}
(f*g)(z, y; \bar y)= \smashoperator{\int_{(u,u',v,v')}} f\left(z+u', y+u; \bar y \right)\ \bar\star\ 
g\left(z-v,y+v+v'; \bar y \right)
\exp({iu_{\al}v^{\al}+iu'_{\al}v'^{\al}})\,,
\ee
where the integration measure over $u, u', v, v'$ is such that $1*1=1$, while $\bar\star$ is \eqref{starproduct} with respect to the variables $\bar y$ only. Explicit formulae for star-product can be derived from \eqref{limst}:
\begin{subequations}\label{star:gen}
\begin{align}
&y_{\al}* =y_{\al}+i\ff{\p}{\p y^{\al}}-i\ff{\p}{\p z^{\al}}\,,\qquad z_{\al}* =z_{\al}+i\ff{\p}{\p
y^{\al}}\,,\\
&* y_{\al}=y_{\al}-i\ff{\p}{\p y^{\al}}-i\ff{\p}{\p z^{\al}}\,,\qquad * z_{\al}=z_{\al}+i\ff{\p}{\p
y^{\al}}\,,\\
&\bar y_{\dal}* =\bar y_{\dal}+i\ff{\p}{\p \bar y^{\dal}}\,,\qquad *
\bar y_{\dal}=\bar y_{\dal}-i\ff{\p}{\p \bar y^{\dal}}\,.
\end{align}
\end{subequations}
The following notation is used for differentiation:
\begin{equation}
    \p_{\alpha} := \frac{\p}{\p y^{\alpha}},\ \bp_{\dot\alpha} := \frac{\p}{\p \y^{\dot\alpha}}.
\end{equation}
In particular, from \eqref{star:gen} one gets the relations
\begin{equation}\label{commutators}
    [y_{\al}, y_{\gb}]_*=2i\gep_{\al\gb}\,,\quad [\bar y_{\dal}, \bar y_{\dgb}]_*=2i\gep_{\dal\dgb}\,,\quad [y_{\al}, z_{\gb}]_*=[\bar y_{\dal}, z_{\gb}]_*=[z_{\al}, z_{\gb}]_*=0\,.
\end{equation}
Note that the $z$-variable lies in the center of the enlarged HS algebra.
We also introduce two types of differential: the space-time one $\dr_x = \dr x^a \frac{\p}{\p x^a}$ and the auxiliary one $\dr_z = \dr z^{\alpha}\frac{\p}{\p z^{\alpha}}$. The new variable allows us to introduce the field $W(x| z; y, \bar y)$, which is designed to capture nonlinear corrections in $\omega$ through its dependence on $z$
\begin{equation}\label{W: decomp}
    W:=\go(y, \bar y)+W_1[\omega, C](z; y, \bar y)+W_2[\omega, C,C](z; y, \bar y)+\dots\ .
\end{equation}
$W$ is a space-time $1$-form which means that the field $\omega$ being also a 1-form enters $W$ strictly linearly meanwhile $C$ can enter $W$ nonlinearly. Hence, we organize the perturbation series in powers of $C$. The equations that generate self-dual HS interactions take the following form: 
\begin{subequations}\label{Eqs}
    \begin{align}   
        &\dr_x W + W*W = 0\,, \label{Eq: dW}\\
        &\dr_z W + \{W, \Lambda\}_* + \dr_x\Lambda=0\,, \label{Eq: dL}\\
        &\dr_x C + \left(W(z; y, \bar y)*_y C - C*_y W(z; -y, \bar y) \right) \Big|_{z= -y} = 0\,. \label{Eq: dC}
    \end{align}
\end{subequations}
Where
\begin{equation}\label{Lambda}
    \Lambda[C]=\dr z^{\al}z_{\al}\int_{0}^{1}d\tau\,\tau\,C(-\tau z,\bar y) e^{i\tau zy}\,.
\end{equation}
And $*_y \equiv \star$ does not affect $z$. 

Equations \eqref{Eqs} are called \textit{generating} in the following sense. The field $W$ should be recovered in powers of $C$ from the second equation \eqref{Eq: dL}. Substitution of this $W$ into \eqref{Eq: dW} yields the unfolded equation \eqref{unfoldedEqs1} with explicit expressions for the vertices $\mathcal{V}$. The same way, the equation \eqref{Eq: dC} gives \eqref{unfoldedEqs2} with explicit expressions for the vertices $\Upsilon$. See \cite{Didenko:2024zpd} for a detailed analysis. The advantage of \eqref{Eqs} is direct access to the local form of the vertices, thanks to a special form of the large star-product \eqref{limst}. We could have used the explicit form of the vertices from \cite{Didenko:2024zpd} in our work, ignoring the auxiliary $z$-dependence, but this method is usually technically more involved. Instead, we will construct the exact solutions of the generating system \eqref{Eqs}, which restores the solution of \eqref{unfoldedEqs} in a linearly-exact form.

\paragraph{Symmetries.} The system \eqref{Eqs} has gauge symmetry
\begin{subequations}\label{symssyms}
\begin{align}
    & \delta_{\xi} W = \dr_x \xi + [W, \xi]_*\,,\\
    & \delta_{\xi} \Lambda = \dr_z \xi + [\Lambda, \xi]_*\,,\\
    & \delta_{\xi} C = \left(\xi(z; y, \bar y) *_y C - C *_y \xi(z; -y, \bar y) \right) \Big|_{z= -y}\, \label{gaugeC}, \\
    & \delta_{\xi} \Lambda = \Lambda[\delta_{\xi}C] \label{gaugeRels}\,.
\end{align}
\end{subequations}
Here $\xi(x| z, y, \y)$ is the gauge parameter, which is $0$-form. Note that the last relation \eqref{gaugeRels} is required for the consistency, that constrains possible $\xi$. These $\xi$ are true local symmetries. One can obtain proper field transformations for $\omega, C$ in terms of \eqref{unfoldedEqs} by performing the procedure of resolving $z$-dependence with the result being \cite{Didenko:2024zpd}
\begin{subequations}
\begin{align}
    & \delta_{\varepsilon} \omega = \dr_x\varepsilon + [\omega, \varepsilon]_* + \sum_{k = 1}^{\infty} \Big(\mathcal{V}(\omega, \varepsilon, C^k) - \mathcal{V}(\varepsilon, \omega, C^k)\Big), \\
    & \delta_{\varepsilon} C = -\varepsilon * C + C * \pi(\varepsilon) - \sum_{k = 2}^{\infty} \Upsilon(\varepsilon, C^k).
\end{align}
\end{subequations}
Here 
\begin{equation}
    \varepsilon := \xi(z, y, \y)\Big|_{z = 0}\,.
\end{equation}
Having fixed a solution, one can ask for global symmetries, which are determined by $\delta_{\xi} W = \delta_{\xi}\Lambda = \delta_{\xi} C = 0$.
\begin{subequations}\label{GlobalSyms}
\begin{align}
    & \dr_x \xi + [W, \xi]_* = 0\,, \label{W0gauge}\\
    & \dr_z \xi + [\Lambda[C], \xi]_* = 0\,, \label{Lambdagauge}\\
    & \left(\xi(z; y, \bar y) *_y C - C *_y \xi(z; -y, \bar y) \right) \Big|_{z= -y} = 0\,. \label{consistencyofLambdagauge}
\end{align}
\end{subequations}
For the solution $(W, C)$ each field implies its own symmetry equation: \eqref{W0gauge} for $W$ and \eqref{Lambdagauge}, \eqref{consistencyofLambdagauge} for $C$. Note that relation \eqref{gaugeRels} is trivially satisfied; however, $\xi$ is still restricted by \eqref{consistencyofLambdagauge}, which becomes a consistency condition for \eqref{Lambdagauge} with respect to $\dr_z$. The solution of \eqref{GlobalSyms} can be constructed perturbatively in the powers of $C$. To do so, one introduces the series
\begin{equation}
    \xi(z, y,\y) = \sum_{k = 0}^{\infty} \xi_k[C^k](z, y, \y) = \xi_0 + \xi_1[C] + \xi_2[C, C] + ... \label{Xi: decomp}
\end{equation}
Note that the perturbative series for $\dr_x C$ contains all powers starting from linear, as it follows from \eqref{Eq: dC}:
\begin{equation}
    \dr_x C = \sum_{k = 0}^{\infty} \Big( C *_y \pi(W_k) - W_k *_y C \Big) \Big|_{z = -y}.
\end{equation}
As a consequence, $\dr_x \xi_k$ contains all powers of $C$ starting from $k$-th:
\begin{equation}
    \dr_x\xi_k[\underbrace{C, ..., C}_{k}] = \xi_k [\dr_x C, \underbrace{C, ..., C}_{k-1}] + (\text{other orderings}) := \sum_{m = k}^{\infty} (\dr_x\xi_k)_m.
\end{equation}
Except for $k = 0$ case, because $\xi_0$ does not depend on $C$. Substitution of the $\xi$ into \eqref{GlobalSyms} gives the following system:
\begin{subequations}\label{GlobalSymsPert}
\begin{align}
    & \dr_x\xi_0 + [\omega, \xi_0]_* = 0, \label{W0gaugePert0}\\
    & \sum_{m = 1}^{k}(\dr_x \xi_m)_{k} + \sum_{m = 0}^k[W_m, \xi_{k-m}]_* = 0,\ k \geqslant 1\,, \label{W0gaugePert}\\
    & \dr_z \xi_0  = 0\,, \label{LambdagaugePert0}\\
    & \dr_z \xi_k + [\Lambda[C], \xi_{k-1}]_* = 0,\ k \geqslant 1\,, \label{LambdagaugePert}\\
    & \left(\xi_k(z; y, \bar y) *_y C - C *_y \xi_k(z; -y, \bar y) \right) \Big|_{z= -y} = 0,\ k \geqslant 0\,. \label{consistencyofLambdagaugePert}
\end{align}
\end{subequations}
Equation \eqref{W0gauge} has been decomposed in powers of $C$ in accordance with \eqref{W: decomp}, \eqref{Xi: decomp}, which gives rise to \eqref{W0gaugePert0} and \eqref{W0gaugePert}. The process of solving \eqref{GlobalSymsPert} is the following. One has to find all $z$-independent $\xi_0$ that satisfy \eqref{W0gaugePert0}, then choose those of them that satisfy the consistency condition \eqref{consistencyofLambdagaugePert}. The next step is solving $\eqref{LambdagaugePert}$ via some homotopy operator\footnote{Equation of the form $\dr_{z}f(z)=\dr z^{\alpha}X_{\alpha}(z)$ can be solved by $f=z^{\alpha}\int_{0}^{1}d\tau X_{\alpha}(\tau z)$, known as standard homotopy. There are different homotopy operators, which differs in $z$-independent cohomology. \label{footnote}} which gives $z$-dependent $\xi_1$. This $\xi_1$ must satisfy \eqref{W0gaugePert} with $k = 1$ and again the consistency condition \eqref{consistencyofLambdagaugePert}. After one finds $\xi_1$, the $\xi_2$ can be restored from \eqref{LambdagaugePert} using homotopy operator. Procedure continues till it finishes.

The gauge transformations form an algebra of local HS symmetries. It is straightforward to check the relations
\begin{equation}
\begin{split}
    & (\delta_{\eta}\delta_{\xi} - \delta_{\xi}\delta_{\eta})W = \delta_{[\xi, \eta]_*}W, \\
    & (\delta_{\eta}\delta_{\xi} - \delta_{\xi}\delta_{\eta})\Lambda = \delta_{[\xi, \eta]_*}\Lambda. \label{lambdacommutator}\\
\end{split}
\end{equation}
The situation is slightly more tricky for $C$ field. The \eqref{gaugeC} itself does not guarantee the algebraic property. One must take into account \eqref{gaugeRels}. Then we get a simple relation $\delta_{\eta}\delta_{\xi}\Lambda = \Lambda[\delta_{\eta}\delta_{\xi}C]$, which entails
\begin{equation}
    \Lambda[(\delta_{\eta}\delta_{\xi} - \delta_{\xi}\delta_{\eta})C] = (\delta_{\eta}\delta_{\xi} - \delta_{\xi}\delta_{\eta})\Lambda = \Lambda[\delta_{[\xi, \eta]_*}C]. \label{lambdacommutatorwithC}
\end{equation}
Notice that $\Lambda[C] = 0 \Leftrightarrow C = 0$. Then the algebraic property follows from \eqref{lambdacommutatorwithC}
\begin{equation}
    (\delta_{\eta}\delta_{\xi} - \delta_{\xi}\delta_{\eta})C = \delta_{[\xi, \eta]_*}C. \label{Ccommutator}
\end{equation}
The solutions of \eqref{GlobalSyms} form an algebra of global higher-spin symmetries. If the generators $\xi, \eta$ satisfy \eqref{GlobalSyms}, then $[\xi, \eta]_*$ satisfies \eqref{GlobalSyms} too, which is due to a discussed relations above. Specifically, in this case the relation \eqref{Ccommutator} is fulfilled simply as an integrability condition for \eqref{lambdacommutator}.

\paragraph{AdS-background.} As was already mentioned, equations \eqref{Eqs} allow for the following vacuum:
\begin{subequations}\label{vac:stnd}
    \begin{align}
        &C=0\,,\\
        &W:=\omega_{AdS}(y, \bar y)=-\frac{i}{4}(\go_{\al\gb}y^\al y^\gb+\bar\go_{\dal\dgb}\bar y^{\dal}\bar y^{\dgb}+2\e_{\al\dgb}y^{\al}\bar y^{\dgb}) \label{AdS}\,,
    \end{align}
\end{subequations}
where 1-forms $\omega^{\alpha\beta}, \bar\omega^{\dot\alpha\dot\beta}$ and $\e^{\alpha\dot\beta}$ are associated respectively with the Lorentz connection and vierbein of the $4d$ AdS. The corresponding Cartan equations following from \eqref{Eq: dW} read
\begin{subequations}\label{AdS:frame}
    \begin{align}
        &\dr_x\go_{\al\gb}+\go_{\al\gga}\wedge\go^{\gga}_{\hspace{5pt}\gb}+ \e_{\al\dgga}\wedge \e_{\gb}^{\hspace{5pt}\dgga}=0\,,\\
        &\dr_x\bar\go_{\dal\dgb}+\bar\go_{\dot{\alpha}\dot{\gamma}}\wedge \bar\go^{\dot{\gamma}}_{\hspace{5pt}\dot{\beta}}+\e_{\gamma\dot{\alpha}}\wedge \e^{\gamma}_{\hspace{5pt}\dot{\beta}}=0\,,\\
        &\dr_x\e_{\al\dal}+\go_{\al\gga}\wedge \e^{\gamma}_{\hspace{5pt}\dal}+\bar\go_{\dal\dgga}\wedge \e_{\alpha}^{\hspace{5pt}\dgga}=0\,.
    \end{align}
\end{subequations}
Here we set the cosmological constant to a number for convenience. The solution \eqref{vac:stnd} describes an empty AdS. In what follows, we use the Poincar\'{e} coordinates
\begin{equation}\label{Poincare}
    ds^2=\frac{1}{r^2}(\dr r^2+\dr\rmx_{\al\gb}\dr\rmx^{\al\gb})\,,
\end{equation}
where $\rmx^{\alpha\beta}$ are the boundary three-dimensional coordinates\footnote{We use $x$ for $4d$ coordinates and $\rmx$ for the $3d$ slice.} 
\begin{equation}
    \rmx_{\al\gb}=\rmx_{\gb\al}\,,
\end{equation}
and the radial coordinate $r$. The background connections \eqref{AdS:frame} can be chosen to be 
\begin{equation}
    \go_{\al\gb}=\frac{i\dr\rmx_{\al\gb}}{2r},\
    \bar\go_{\dal\dgb}=-\frac{i\dr\rmx_{\dal\dgb}}{2r},\
    \e_{\al\dgb}=\frac{i\epsilon_{\al\dgb}\,\dr r - \dr\rmx_{\al\dgb}}{2r}\,. \label{vierbein}
\end{equation}
Throughout the paper we do not  differentiate dotted and undotted indices converting one to another via $\gep_{\dgb\al}$ according to the following convention:
\begin{align}
   &\bar y_{\al}:=\bar y^{\dgb}\gep_{\dgb\al}\,,\qquad y_{\dal}:=y^{\gb}\gep_{\gb\dal}\,.\\ 
   &\rmx_{\al\dgb}:=\rmx_{\al}{}^{\gb}\gep_{\gb\dgb}\,,\qquad \rmx_{\dal\dgb}=\rmx^{\al\gb}\gep_{\al\dal}\gep_{\gb\dgb}\,.
\end{align}
It is convenient to change variables
\begin{equation}\label{y:pm}
    y^\pm_{\al}:=y_{\al}\pm i\bar y_{\al}\,,\qquad [y^\pm_{\al}, y^\pm_{\gb}]_*=0\,,
\end{equation}
Star-product rules change correspondingly
\begin{equation}
\begin{split}
    & y^{\pm}_{\alpha} * = y^{\pm}_{\alpha} + 2i\p^{\mp}_{\alpha} - i\frac{\p}{\p z^{\alpha}},\ * y^{\pm}_{\alpha} = y^{\pm}_{\alpha} - 2i\p^{\mp}_{\alpha} - i\frac{\p}{\p z^{\alpha}}\,, \\
    & \p^{\pm}_{\alpha} := \frac{\p_{\alpha} \mp i\bp_{\alpha}}{2},\ \p^{\pm}_{\alpha} y^{\pm}_{\beta} = \epsilon_{\alpha\beta}\,.
\end{split}
\end{equation}
For an empty AdS vacuum \eqref{vac:stnd} one can find global symmetries. Equations \eqref{GlobalSyms} simplifies down to
\begin{equation}
    \dr_x \xi + [\omega_{AdS}, \xi]_* = 0\,. \label{AdSGlobalSyms}
\end{equation}
It is quite easy to solve \eqref{AdSGlobalSyms} using $y^{\pm}$ variables. The general solution $\xi(x| y^+, y^-)$ is given by the following polynomials:
\begin{equation}
\begin{split}
    &\xi_{n, m} := \frac{\xi^{\mu(n-m),\nu(m)}}{(n-m)!m!} y^+_{\mu(n-m)}y^-_{\nu(m)} + ... + \frac{\xi^{\mu(n)}}{n!} y^+_{\mu(n)} = \sum_{k = 0}^m \frac{\xi^{\mu(n-k),\nu(k)}}{(n-k)!k!} y^+_{\mu(n-k)}y^-_{\nu(k)}, \\
    & \xi^{\mu(n-k),\nu(k)} = r^{k - \frac{n}{2}} \sum_{l = 0}^{m-k} \begin{pmatrix}
        n-k \\
        l
    \end{pmatrix} \Big(-\frac{3i}{4} \Big)^{l} \underbrace{\rmx^{\mu}{}_{\nu}...\rmx^{\mu}{}_{\nu}}_{l} \eta^{\mu(n - k - l), \nu(k + l)}, \\
    & \eta^{\mu(n - k - l), \nu(k + l)} = \text{const},\ 
    \begin{pmatrix}
        n-k \\
        l
    \end{pmatrix} = \frac{(n-k)!}{l!(n-k-l)!}\ \text{is the number of combinations.}
\end{split} \label{AdSGenerators}
\end{equation}
Specifically, for $n = 2$ we get the spin-$2$ algebra in accordance with \eqref{Omega:phys}:
\begin{equation}
\begin{split}
    & \xi_{2, 0} = \frac{\eta^{\mu(2)}}{2r}y^+_{\mu}y^+_{\mu},\ \xi_{2,1} = \eta^{\mu,\nu}y^+_{\mu}y^-_{\nu} + \frac{1}{2r}\Big(\eta^{\mu(2)} - \frac{3i}{2}\rmx^{\mu}{}_{\nu}\eta^{\mu,\nu}\Big)y^+_{\mu}y^+_{\mu}\,, \\
    & \xi_{2,2} = \frac{r}{2} \eta^{\nu(2)}y^-_{\nu}y^-_{\nu} + \Big(\eta^{\mu,\nu} - \frac{3i}{4}\rmx^{\mu}{}_{\nu}\eta^{\nu(2)}\Big)y^+_{\mu}y^-_{\nu} + \frac{1}{2r}\Big(\eta^{\mu(2)} - \frac{3i}{2}\rmx^{\mu}{}_{\nu}\eta^{\mu,\nu} - \frac{9}{16} \rmx^{\mu}{}_{\nu}\rmx^{\mu}{}_{\nu}\eta^{\nu(2)} \Big) y^+_{\mu}y^+_{\mu}\,.
\end{split}
\end{equation}
The $\xi_{2,0}$ has $3$ independent parameters; $\xi_{2,1}$ has 3 parameters for symmetric part of $\eta^{\mu,\nu}$ and $1$ additional for the trace part $\eta^{\mu,\nu} \sim \epsilon^{\mu\nu}$; $\xi_{2,2}$ has also $3$ parameters. Together spin-$2$ algebra admits $10$ independent parameters. 
We gain nothing but the algebra of isometries of an AdS$_4$. Explicitly, the $so(3,2)$ can be written in a form of $3d$ conformal generators:
\begin{subequations}\label{confal}
    \begin{align}
        &P_{\al\gb}=iy^+_{\al}y^{+}_{\gb}\,,\\
        &L_{\al\gb}=i(y^+_{\al}y^-_{\gb}+y^+_{\gb}y^-_{\al})\,,\\
        &K_{\al\gb}=iy^-_{\al}y^{-}_{\gb}\,,\\
        &D=\frac{i}{8}y^{-}_{\al}\bar y^{+\al}=-\frac{1}{4}y\bar y\,,\label{D}
    \end{align}
\end{subequations}
where $P$, $L$, $K$ and $D$ are the generators of translations, Lorentz boosts, special conformal transformations, and dilatation, respectively. They enter $\xi_2$ generators according to the scheme below
\begin{equation}
\begin{split}
    & \xi_{2, 0} \longleftrightarrow P,\quad \text{symmetric part of}\ \xi_{2,1} \longleftrightarrow L, \\
    & \text{trace of}\ \xi_{2,1}\ \longleftrightarrow D,\quad  \xi_{2,2} \longleftrightarrow K.
\end{split}
\end{equation}
The commutation relations read
\begin{subequations}   
    \begin{align}
        &[P_{\alpha\beta}, L_{\mu\nu}] = -4(\epsilon_{\alpha\nu}P_{\mu\beta} + \epsilon_{\beta\nu}P_{\mu\alpha} + \epsilon_{\alpha\mu}P_{\nu\beta} + \epsilon_{\beta\mu}P_{\nu\alpha})\,,\\
        &[L_{\alpha\beta},L_{\mu\nu}] = 4(\epsilon_{\alpha\mu}L_{\beta\nu} + \epsilon_{\beta\mu}L_{\alpha\nu} + \epsilon_{\alpha\nu}L_{\beta\mu} + \epsilon_{\beta\nu}L_{\alpha\mu})\,,\\
        &[K_{\alpha\beta}, P_{\mu\nu}] = -2(\epsilon_{\alpha\mu}L_{\beta\nu} + \epsilon_{\alpha\nu}L_{\beta\mu} + \epsilon_{\beta\mu}L_{\alpha\nu} + \epsilon_{\beta\nu}L_{\alpha\mu}) - 32(\epsilon_{\alpha\mu}\epsilon_{\beta\nu} + \epsilon_{\alpha\nu}\epsilon_{\beta\mu})D\,,\\
        &[D, P_{\alpha\beta}] = - P_{\alpha\beta}\,,\qquad [D, K_{\alpha\beta}] =  K_{\alpha\beta}\,, \label{confdilattrans}\\
        &[D, L] = [P,P] = [D,D] = [K,K] = 0\,.
    \end{align}
\end{subequations}
Vacuum connection then expresses as follows:
\begin{equation}\label{Omega:vac}
    \omega_{AdS} = \ff{\dr\rmx^{\al\gb}}{8i r}P_{\alpha\beta} - \ff{\dr r}{r}D\,.
\end{equation} 
Translations and Lorentz boosts separately form a $3d$ Poincar\'{e} algebra, which is our interest for the next section. We will construct solutions that has a $2+1$ Poincar\'{e} as a leftover global space-time symmetry.

\section{Poincar\'{e} invariance}
In this section, we focus on solutions which leave the background to be an AdS, meanwhile non-zero $C$ must preserve $3d$ Poincar\'{e} symmetry. Specifically, we are interested in solutions of the form
\begin{equation}
\begin{split}
    & W(z, y, \y) = \omega_{AdS} + W_1[C](z, y, \y) + W_2[C^2](z, y, \y) + ...\, \\
    & W(z, y, \y)|_{z = 0} = \omega_{AdS}\,.
\end{split} 
\end{equation}
Global symmetries of that solution have to contain Poincar\'{e} generators
\begin{equation}\label{poincare}
    \varepsilon_P = \varepsilon_0^{\alpha\beta} P_{\alpha\beta}\,,\ \varepsilon_L = \varepsilon_1^{\alpha\beta} L_{\alpha\beta} + \varepsilon_2^{\alpha\beta} P_{\alpha\beta}\,.
\end{equation}
This is the starting point for perturbative analysis. In terms of \eqref{GlobalSymsPert} we have chosen the $\xi_0$ to be either $\varepsilon_P$ or $\varepsilon_L$. In the end, this generators can get a $z$-dependent corrections. Still $z$-independent part has to remain \eqref{poincare} in order to be Poincar\'{e}. This can be seen from  \eqref{W0gaugePert0}
\begin{equation}
    \dr_x\xi_0 + [\omega_{AdS}, \xi_0] = 0\,. \label{zindepGlobSyms1}
\end{equation}
This equation is identical to one that arises for an empty AdS; hence, its solutions are given via \eqref{AdSGenerators}, which contain \eqref{poincare} as a subalgebra.

The matter of our interest now is the consistency condition \eqref{consistencyofLambdagaugePert}:
\begin{equation}
    \xi_0 * C - C * \pi(\xi_0) = 0\,. \label{zindepGlobSyms3}
\end{equation}
It turns out that this equation significantly constraints $C$. For each generator we have:
\begin{subequations} \label{PoincareInvariance}
\begin{align}
    & \varepsilon^{\alpha\beta}\Big(P_{\alpha\beta} * C - C * \pi(P_{\alpha\beta})\Big) = 0\,, \label{TranslationalInvariance}\\
    & \varepsilon^{\alpha\beta}\Big(L_{\alpha\beta} * C - C * \pi(L_{\alpha\beta})\Big) = 0\,, \label{LorentzInvariance}
\end{align}
\end{subequations}
where the parameter $\varepsilon^{\alpha\beta}$ is an arbitrary symmetric tensor. First one \eqref{TranslationalInvariance} tells that $C$ has to be $\rmx$-independent\footnote{Here $\rmx$ is a boundary coordinate, which should not be confused with $x = (r, \rmx)$}. To see this, recall the equation imposed on $C$ \eqref{Eq: dC}. It splits into two along $\dr\rmx^{\alpha\beta}$ and $\dr r$
\begin{equation}
\begin{split}
    & \dr_{\rmx}C + \frac{\dr\rmx^{\alpha\beta}}{8r}\Big(P_{\alpha\beta} * C - C * \pi(P_{\alpha\beta})\Big) = 0\,, \\
    & \dr_{r}C - \frac{\dr r}{r}\{D,C\} = 0\,. \label{eq:dr}
\end{split}
\end{equation}
Comparing \eqref{TranslationalInvariance} and \eqref{eq:dr}, we conclude that $\dr_{\rmx} C = 0$; \eqref{Eq: dC} restores $r$-dependence. Second restriction \eqref{LorentzInvariance} needs to be calculated:
\begin{equation}
    \varepsilon^{\alpha\beta}\Big( L_{\alpha\beta} * C - C * \pi(L_{\alpha\beta}) \Big) = -8\varepsilon^{\alpha\beta}(y_{\alpha}\p_{\beta} + \y_{\alpha}\bp_{\beta})C\,.
\end{equation}
Using the general form of $C$,
\begin{equation}
    C(y, \y) = \sum_{m,n = 0}^{\infty} \frac{C^{\mu(n), \nu(m)}}{n!m!} y_{\mu(n)} \y_{\nu(m)}\,,
\end{equation}
one obtains the equation imposed on components:
\begin{equation}
    \varepsilon^{\mu}{}_{\alpha}C^{\alpha\mu(n-1), \nu(m)} + \varepsilon^{\nu}{}_{\alpha}C^{\mu(n), \alpha\nu(m-1)} = 0\,. \label{Cconstraint}
\end{equation}
This equation has to be satisfied for an arbitrary symmetric $\varepsilon^{\alpha\beta}$. Vanishing must occur only due to the index symmetries of $C^{\mu(n), \nu(m)}$. One should decompose tensor $C^{\mu(n), \nu(m)}$ into components of different symmetry types, which are encoded by Young diagrams. Trivial vanishing occurs only for the totally antisymmetric component with $m = n$
\begin{equation}
    C^{\mu(n), \nu(n)} = C(r) \cdot \underbrace{\epsilon^{\mu\nu}...\epsilon^{\mu\nu}}_{n\ \text{times}}\,.
\end{equation}
And for any other Young diagram one gets the algebraic constraint on $\varepsilon^{\alpha\beta}$. Hence, we conclude that the only way to satisfy \eqref{Cconstraint} trivially is to set
\begin{equation}
    C^{\mu(n), \nu(m)} = 0,\ \text{if}\ m \neq n\,.
\end{equation}
We see that $C$ depends on $\epsilon^{\alpha\beta}y_{\alpha}\y_{\beta} \equiv y\y$ only. In other words, the Poincar\'e-invariance constrains the spectrum of the vacuum fields, demanding the solution to be scalar. Eventually, \eqref{PoincareInvariance} leaves us with the only possible ansatz:
\begin{equation}
    C = C(r, y\y).
\end{equation}
Substitution into equations gives the following two-parametric solution:
\begin{equation}\label{scalar}
    \begin{split}
        & C_0 = \nu_1\cdot r \cdot e^{y\y} + \nu_2\cdot r^2 \cdot (1 + y\y) e^{y\y},\ \nu_1, \nu_2 = \text{const}\,, \\
        & W_0 = \omega_{AdS} + \nu_2\cdot \frac{i r}{2}\dr \rmx^{\al\gb}z_{\al}z_{\gb}\int_0^1d\tau\tau (1-\tau)e^{i\tau zy^+}\,.
    \end{split}
\end{equation}
This solution was obtained earlier in \cite{Didenko:2025xca}, see Appendix A in this reference for a detailed analysis. 

\section{New vacuums}
New exact solutions arise when one relaxes the leftover Poincar\'{e} symmetry condition. We still look for the solutions on AdS background, i.e. $W|_{z = 0} = \omega_{AdS}$. In what follows, we solve the system \eqref{Eqs}. We are concerned with the lowest perturbation order in $C$, because the solutions actually does not obtain $z$-dependent corrections, as will be shown below.

\paragraph{Equation on $C$.} The equation imposed on the $C$-module is \eqref{Eq: dC} with vanishing r.h.s., where in the lowest order we substitute $W \rightarrow \omega_{AdS}$
\begin{equation}
    \dr_x C + \omega_{AdS}*C - C*\pi(\omega_{AdS}) = 0\,.
\end{equation}
To analyze it, we use the following ansatz generalizing the scalar solution \eqref{scalar}
\begin{equation}
    C_{n,m} := r^{(n+2)/2}\cdot C^{\mu(n),\nu(m)} y_{\mu(n)}\y_{\nu(m)} \cdot e^{y\y}\,.
\end{equation}
Direct substitution gives two new possible solutions. They are the following:
\begin{subequations}
\begin{align}
    & C_{n,0} = r^{(n+2)/2} C^{\alpha(n)}\cdot y_{\alpha(n)}\cdot e^{y\y},\  C^{\alpha(n)} = \text{const}\,,\label{holSol}\\
    & C_{0,m} = r^{(m+2)/2} \bar C^{\beta(m)} \cdot \y_{\beta(m)}\cdot e^{y\y}, \bar C^{\alpha(n)} = \text{const}\,. \label{aholSol}
\end{align}
\end{subequations}
We refer to \eqref{holSol} and \eqref{aholSol} as \textit{first and second branches} respectively. Each branch carries the spin-$S$ degrees of freedom: 
\begin{equation}
    s = \frac{n}{2}\ \text{for}\ C_{n, 0}\quad \text{and}\quad s = -\frac{m}{2}\ \text{for}\ C_{0, m}\,.
\end{equation}
Recall that $S$ is the \textit{spin} and $s$ is the \textit{helicity}. One can expect that only one branch should be a solution of the full system \eqref{Eqs} due to the self-duality. This is indeed the case, as we will see below.

\paragraph{Equation on $\Lambda$.} Equation \eqref{Eq: dL} in the lowest order is as follows:
\begin{equation}
    \dr_x \Lambda + \{\omega_{AdS}, \Lambda\} = 0\,. \label{Lambdaeq}
\end{equation}
Expressions for $\Lambda$ for each branch reveals the difference
\begin{subequations}
\begin{align}
    & \Lambda[C_{n,0}] = (-)^{n}r^{(n+2)/2} C^{\alpha(n)} \dr z^{\mu}\int_0^1\dr\tau\ \tau^{n+1} z_{\mu}z_{\alpha(n)}e^{i\tau zy^+}\,, \label{Ln}\\
    & \Lambda[C_{0,m}] = r^{(m+2)/2} C^{\beta(m)} \dr z^{\mu}\int_0^1\dr \tau\ \tau z_{\mu}\y_{\beta(m)}e^{i\tau zy^+}\,. \label{Lm}
\end{align}
\end{subequations}
Recall the definition of the $\Lambda$ field \eqref{Lambda}. Note that $\Lambda$ changes $y$ to $z$, which is crucial as $z$ commutes with everything \eqref{commutators}. The second branch cancels due to this fact. Specifically, translations commute with the first branch and do not commute with the second one
\begin{equation}\label{commWithTransls}
    [P_{\alpha\beta}, \Lambda[C_{n, 0}]] = 0,\ [P_{\alpha\beta}, \Lambda[C_{0, m}]]\neq 0\,.
\end{equation}
The equation \eqref{Lambdaeq} takes the form
\begin{subequations}
\begin{align}
    & \dr_{r} \Lambda - \frac{\dr r}{r}[D, \Lambda]  = 0\,, \\
    & \frac{\dr\rmx^{\alpha\beta}}{8ir}[P_{\alpha\beta}, \Lambda] = 0\,. \label{EqdLsec} 
\end{align}
\end{subequations}
The second branch does not satisfy \eqref{EqdLsec}. The first branch indeed is a solution. This result relies on \eqref{commWithTransls} and an observation
\begin{equation}
    [D, \Lambda[C_{n, 0}]] = \frac{n+2}{2} \Lambda[C_{n, 0}]\,.
\end{equation}

\paragraph{The second branch is not a solution.}
One can expect that $z$-dependent corrections to $\omega_{AdS}$ could turn the second branch into a solution. Let us stress again that $W|_{z = 0} = \omega_{AdS}$. We now show that perturbative process is doomed for the second branch. Recall expansion for $W$
\begin{equation}
    W = \omega_{AdS} + W_1[\omega_{AdS}, C] + W_2[\omega_{AdS}, C, C] + ...,\ C = r^{(m+2)/2} \bar C^{\beta(m)} \cdot \y_{\beta(m)} \cdot e^{y\y} \,.
\end{equation}
Substitution into \eqref{Eqs} gives a chain of equations for each $W_k$. Let us look at the first order
\begin{subequations} \label{linearIMmprovement}
\begin{align}
    & \dr_{x} W_1 + \{\omega_{AdS}, W_1\}_* = 0\,, \\
    & \dr_z W_1 + \{\omega_{AdS}, \Lambda[C]\}_* + \Lambda\Big[\omega_{AdS} *_y C - C *_y \pi(\omega_{AdS})\Big] = 0\,, \label{ahahah} \\
    & \Big(W_1 *_y C - C *_y \pi(W_1)\Big)\Big|_{z = -y} = 0\,. \label{thirdCondonImprvmnt}
\end{align}
\end{subequations}
Equation \eqref{ahahah} can be solved using standard homotopy operator:
\begin{equation}
    W_1 = -\Delta_0 \{\omega_{AdS}, \Lambda[C]\} = im\cdot r^{(m+2)/2} \bar C^{\beta(m)}\cdot \bold e^{\gamma}{}_{\beta} z_{\gamma}\y_{\beta(m-1)} \int_0^1 \dr\tau(1-\tau) e^{i\tau zy^+}\,.
\end{equation}
The first equality uses the fact that $\Delta_0 \Lambda[C] = 0$ for arbitrary $C$. Next step is to check \eqref{thirdCondonImprvmnt}. Straightforward calculation shows that it is not satisfied
\begin{equation}
    \Big(W_1 *_y C - C *_y \pi(W_1)\Big)\Big|_{z = -y} \neq 0\,.
\end{equation}
This result shows that the second branch cannot be a solution on the AdS background. Still one can try to look for the proper $z$-independent cohomology in order to make the second branch a solution on a different background. This is an interesting question, which we leave for future investigations.

\paragraph{Scalar second branch.} Let us show how to restore scalar branches. The first scalar branch is obviously $C_{0,0}$. Recall the equations imposed on $C$
\begin{subequations}
\begin{align}
    & \dr_{\rmx}C + \frac{i\dr\rmx^{\alpha\beta}}{2r}(y_{\alpha}\p_{\beta} - \y_{\alpha}\bp_{\beta} + y_{\alpha}\y_{\beta} - \p_{\alpha}\bp_{\beta})C = 0\,, \label{eq:dx1}\\
    & \dr_{r}C + \frac{\dr r}{2r}(y\y - \p\bp) C = 0\,.\label{eq:dr1}
\end{align}
\end{subequations}
Equation \eqref{eq:dx1} admits a solution $C_{1,1}$ (i.e., for $m = n = 1$) with antisymmetric coefficient $C^{\alpha\beta} \sim \epsilon^{\alpha\beta}$
\begin{equation}
    C_{1,1} = C_{11}(r) \cdot y\y \cdot e^{y\y}\,.
\end{equation}
Though $C_{1,1}$ does not satisfy \eqref{eq:dr1}:
\begin{equation}
    \p_{r}C_{11}\cdot y\y\cdot e^{y\y} \neq \frac{1}{r}\Big(1 + 2 y\y\Big) C_{11} e^{y\y}\,.
\end{equation}
We can combine the two solutions found of \eqref{eq:dx1} in order to construct a solution of \eqref{eq:dr1}:
\begin{equation}
\begin{split}
    & C_2 := C_{0,0} + C_{1,1} = C_{00}(r)\cdot e^{y\y} + C_{11}(r)\cdot y\y \cdot e^{y\y}\,, \\
    & \p_{r} C_2 = \frac{C_{00}e^{y\y}}{r} + \frac{C_{11} e^{y\y}}{r}(1 + 2 y\y)\,, \\
    & \Rightarrow C_2 = \nu_2\cdot r^2\cdot (1 + y\y) \cdot e^{y\y}\,.
\end{split}
\end{equation}
There is an important difference between the scalar field and the second branch $C_{0,m}$. The second scalar branch $C_2$ needs z-dependent corrections to the connection as it was proposed for the $C_{0,m}$. Equations \eqref{linearIMmprovement} have to be satisfied for $C_2$. The crucial difference comes from \eqref{thirdCondonImprvmnt}. The scalar second branch carries a $y\y$ contraction which becomes $yz$ after computing \eqref{thirdCondonImprvmnt} and vanishes trivially because of the projection $z = -y$. This cannot happen for the $C_{0,m}$.

\paragraph{Summary.}
A family of solutions has been obtained
\begin{equation}
\begin{split}
    & W = \omega_{AdS}\,, \\
    & C_{n} = r^{\frac{n+2}{2}} C^{\alpha(n)}\cdot y_{\alpha(n)} e^{y\y},\ C^{\alpha(n)} = \text{const}\,.
\end{split}
\end{equation}
\begin{itemize}
\item $C_n$ carries a helicity-$\frac{n}{2}$ degrees of freedom. The dilatation operator $D $ acts on $C_n$ in a twisted representation, while $\Lambda[C_n]$ lies in an adjoint one:
\begin{equation}
    \{D, C_n\}_* = \frac{n+2}{2} C_n,\ [D, \Lambda[C_n]]_* = \frac{n+2}{2}\Lambda[C_n]\,.    
\end{equation}
$C_n$ has a conformal weight $\Delta_n = \frac{n+2}{2}$.
    
\item Solution $C_n$ satisfies an unfolded equation:
    \begin{equation}
        \dr_x C_n + \omega_{AdS} * C_n - C_n * \pi(\omega_{AdS}) = 0\,.
        \label{unfoldedCurrents}
    \end{equation}
    Namely, all the vertices vanish:
    \begin{equation}
    \begin{split}
        & \mathcal{V}_n(\omega_{AdS}, \omega_{AdS}, C^k_{n}) = 0\,, \\
        & \Upsilon_n(\omega_{AdS}, C^k_{n}) = 0\,.
    \end{split}
    \end{equation}
    This is straightforward to check using the explicit expressions for vertices, which can be found in \cite{Didenko:2024zpd}.

\item Unfolded equation \eqref{unfoldedCurrents} corresponds to a so-called \textit{rank-2} type of equations \cite{Gelfond:2003vh}. Following \cite{Vasiliev:2012vf}, one can use the \textit{$T$-module} ansatz:
\begin{equation}
    C = r \cdot e^{y\y} \cdot T(w, \bar w| x),\quad w = \sqrt{r} y,\ \bar w = \sqrt{r} \y\,.
\end{equation}
Substituting this into \eqref{unfoldedCurrents}, one gets
\begin{subequations}
\begin{align}
    & \dr_x T - \frac{i}{2} \dr\rmx^{\alpha\beta} \p_{\alpha}\bp_{\beta}T = 0\,, \label{currentsconserveddx}\\
    & \p_r T - \frac{1}{2}\p\bp T = 0\,, \label{currentsconserveddr}
\end{align}
\end{subequations}
where $\p, \bp$ differentiate first and second argument of $T(w, \bar w)$ respectively. Equation \eqref{currentsconserveddx} describes an off-shell set of conserved currents on the $3d$ boundary. Equation \eqref{currentsconserveddr} gives the evolution from slice to slice along the radial coordinate in the bulk. The system describes the following set of primary currents
\begin{equation}
    J = T(w, 0),\ \bar J = T(0, \bar w),\ J^{asym} = w\bar w\cdot \p\bp T(0, 0)\,.
\end{equation}
All the other components of $T$ are derivatives of the primaries. Obtained solutions are nothing but $J$:
\begin{equation}
\begin{split}
    & r \cdot e^{y\y} \cdot J = \sum_{n = 0}^{\infty} C_n\,, \\
    & J = \sum_{n = 0}^{\infty} (\sqrt{r})^{n} y_{\alpha(n)} C^{\alpha(n)},\ C^{\alpha(n)} = \text{const}\,.
\end{split}
\end{equation}
Currents $\bar J$ are absent, which is due to the self-duality of the system. The second scalar branch corresponds to $J^{asym}$. These currents conserve trivially as they do not depend on the boundary coordinate $\rmx$:
\begin{equation}
    \frac{\p}{\p\rmx^{\alpha\alpha}} J^{\alpha(2S)} = 0\,.
\end{equation}
\end{itemize}

\section{Discussion}
Exact solutions in HS theory are rare; see \cite{Sezgin:2005pv, Sezgin:2005hf, Iazeolla:2007wt, Didenko:2009td, Iazeolla:2011cb, Iazeolla:2015tca, Iazeolla:2017dxc, Sundell:2016mxc, Aros:2017ror}. In the self-dual case, the situation is simpler: some exact solutions are linearly exact \cite{Didenko:2025xca} and even simpler in the flat case when the cosmological constant is zero; see, e.g., \cite{Skvortsov:2024rng, Tran:2025yzd}. 
In this paper we obtained a family of new vacuums in the $4d$ self-dual HS theory. The family contains a single branch for each spin except for the scalar, which has two branches, meanwhile space-time geometry is an AdS. The solution of spin $S$ has the conformal dimension $\Delta_S = S+1$, while the scalar has two branches of the dimensions $\Delta_{0,1} = 1$ and $\Delta_{0,2} = 2$. 
The scalar $S = 0$ and a fermion $S = 1/2$ can be thought of as massless fields in the bulk:
\begin{equation}
	\phi(r) = \nu_1 \cdot r + \nu_2\cdot r^2,\ \psi_{\alpha} = C_{\alpha} \cdot r^{3/2},\quad \nu_1, \nu_2, C_{\alpha} = \text{const}.  
\end{equation}
They depend on the Poincar\'{e} radial coordinate $r$ only. The remaining fields do not have corresponding gauge fields and are represented solely by the Weyl tensors. A proper interpretation comes from the $AdS/CFT$ dictionary. On the $3d$ conformal boundary the obtained fields correspond to a set of the off-shell conserved currents. The spin-$S$ current is of the conformal dimension $\Delta_S = S + 1$ 
\begin{equation}
	J_S = r^{S} \cdot C^{\alpha(2S)} \cdot y_{\alpha(2S)},\ \{D, J_S\} = \Delta_S \cdot J_S,
\end{equation}
which corresponds to a unitarity bound in $CFT_3$. These currents are trivial as they do not depend on the boundary coordinate $\rmx$ at all.

It has been shown that scalar branches are the only solutions implementing the symmetry breaking from AdS$_4$ isometries down to a $3d$ Poincar\'{e}. Global symmetries of the scalar branches contain Poincar\'{e} algebra. To find all the generators one has to solve global symmetry equations \eqref{GlobalSyms}. Scalar solutions provide a toy model for studying symmetry breaking in HS theories. One example exploiting the scalar branch was examined recently in \cite{Didenko:2025xca}. Let us note two crucial things about that analysis. First, it was only the first branch that was used to get the free dynamics. Obviously, one should do the same for the second branch which can provide some new results. Second, the free dynamics has been constructed for the gauge field of the special form $\bold w(y^+)$. General case remains to be studied. 

\section*{Acknowledgments}
The author would like to thank V. Didenko for fruitful discussions and D. Valerev for his comments on the draft.
    

\bibliographystyle{plain}

\bibliography{references}
   
\end{document}